# Large Fermi Surface Expansion through Anisotropic *c-f* Mixing in the Semimetallic Kondo Lattice System CeBi

Peng Li[1], Zhongzheng Wu[1], Fan Wu[1], Chunyu Guo[1], Yi Liu[2], Haijiang Liu[3], Zhe Sun[2], Ming Shi[3], Fanny Rodolakis[4], Jessica L McChesney[4], Chao Cao[5], Frank Steglich[1*], Huiqiu Yuan[1,6,7*], and Yang Liu[1,6,7*]

[1]Center for Correlated Matter and Department of Physics, Zhejiang University, Hangzhou, P. R. China

[2]National Synchrotron Radiation Laboratory, University of Science and Technology of China, Hefei, P. R. China

[3]Paul Scherrer Institute, Swiss Light Source, CH-5232 Villigen PSI, Switzerland

[4]Advanced Photon Source, Argonne National Lab, 9700 South Cass Avenue, Argonne, Illinois 60439, USA

[5]Department of Physics, Hangzhou Normal University, Hangzhou, P. R. China

[6]Zhejiang Province Key Laboratory of Quantum Technology and Device, Zhejiang University, Hangzhou, P. R. China

[7]Collaborative Innovation Center of Advanced Microstructures, Nanjing University, Nanjing, China

*Corresponding authors: yangliuphys@zju.edu.cn, hqyuan@zju.edu.cn, Frank.Steglich@cpfs.mpg.de

## Abstract

Using angle-resolved photoemission spectroscopy (ARPES) and resonant ARPES, we report evidence of strong anisotropic conduction-$f$ electron mixing ($c$-$f$ mixing) in CeBi by observing a largely expanded Ce-$5d$ pocket at low temperature, with no change in the Bi-$6p$ bands. The Fermi surface (FS) expansion is accompanied by a pronounced spectral weight transfer from the local $4f^{\,0}$ peak of Ce (corresponding to $Ce^{3+}$) to the itinerant conduction bands near the Fermi level. Careful analysis suggests that the observed large FS change (with a volume expansion of the electron pocket up to 40%) can most naturally be explained by a small valence change (~ 1%) of Ce, which coexists with a very weak Kondo screening. Our work therefore provides evidence for a FS change driven by real charge fluctuations deep in the Kondo limit, which is made possible by the low carrier density.

Intermetallic compounds of some rare earths (RE, notably Ce and Yb) frequently behave as Kondo-lattice (KL) systems. Well below the Kondo temperature $T_K$, they can develop different types of interesting ground-state properties, including heavy Fermi-liquid, unconventional superconducting, magnetically ordered, or Kondo insulating/semi-metallic phases [1,2,3,4,5]. For these materials, $T_K$ may vary between a few K ("Kondo limit") and a few hundred K ("Intermediate Valence (IV) limit"). In the Kondo limit, Kondo screening of the local 4$f$-moments plays the key role over a wide temperature range up to above $T_K$. This is a many-body process involving the local 4$f$-electron states and electron-hole pairs at the Fermi level ($E_F$), leading to spin flips without changing the 4$f$-occupation ("virtual charge fluctuations"). Meanwhile, there exists a minority of "real charge (valence) fluctuations" due to the weak (covalent) mixing or hybridization, which is necessary for the Kondo screening to operate, as described by the periodic Anderson model [6]. Therefore, for KL systems, the Ce valence is not exactly integer (typically 3< ν<3.1) and increases slightly upon cooling. In the IV limit, where ν is typically between 3.2 and 3.5, the valence fluctuations predominate, but the Kondo spin flips are still operating. This is illustrated by the negative temperature coefficient (NTC) of the resistivity, $\rho(T) \sim -\log T/T_K$, in the canonical IV compound CePd$_3$ [7]. On the other hand, it was found for dilute IV Eu ions in ScAl$_2$ in the absence of Kondo spin flips that the incremental Eu-derived resistivity exhibits a positive $T$-coefficient [8], while a Kondo-type NTC is observed for a few Eu-based compounds [9]. It is therefore a big challenge to resolve valence fluctuations from Kondo spin flips, particularly for KL systems with low $T_K$, where the Kondo screening is predominant, and the effect of tiny valence fluctuations is almost impossible to identify unambiguously in transport and thermodynamic measurements.

As will be shown below, low carrier density KL systems provide an interesting case to resolve valence-fluctuation-induced changes in the Fermi surface (FS), deep in the Kondo limit. The low carrier density significantly reduces the Kondo screening, while the small FS allows for easy detection of any change in the Fermi wavevector $k_F$ due to valence fluctuations. Low carrier density Kondo systems have attracted considerable interest recently, in view of novel phenomena such as unconventional quantum criticality and correlated topological states [10,11,12,13,14]. This has motivated us to study CeBi, a well-known low carrier density KL system (~0.03 e$^-$/Ce [15]). CeBi was extensively studied in the past because of its anomalous electromagnetic properties [16,17,18,19,20,21]. Its resistivity shows NTC suggesting Kondo screening, with a very low $T_K$ due to the few carriers available. As a consequence, the 4$f$-moments are only very weakly screened and undergo antiferromagnetic (AFM) order at $T_{N1} \simeq 25$ K and $T_{N2} \simeq 12$ K, respectively [16,22]. According to Kasuya and collaborators [17,23], the weak Kondo interaction becomes substantially strengthened by the strong mixing between the Bi-6$p$ and Ce-4$f$ states ($p$-$f$ mixing). An alternative proposal, however, has argued for non-Kondo exchange to explain the electromagnetic properties [18].

Bulk CeBi crystallizes in the simple rock-salt structure; its three-dimensional (3D) Brillouin zone (BZ) and associated two-dimensional (2D) BZ are shown in Fig. 1(a). At high temperature, the Ce has a valence close to +3 (with one 4$f$ electron), and the FS consists of two small hole pockets at the $\Gamma$ point (from Bi-6$p$ orbitals) and one small electron pocket at each symmetry-equivalent $X$ point (from Ce-5$d$ orbitals) [14,24]. The bottom panel in Fig. 1(a) displays the 3D bulk FS from density-functional theory (DFT) calculations, assuming the 4$f$ electrons to be completely localized [24,25]. The experimental FSs near the 2D $\overline{\Gamma}$ and $\overline{M}$ points (the $X$ point in 3D bulk BZ projects onto the $\overline{M}$ point in 2D BZ), obtained by ARPES measurements at both 10 K and ~37 K, are displayed in Fig. 1(b,c), in comparison with the $k_z$ projected DFT calculations. Good overall agreements between the experiments and calculations can be found, implying that the 4f electrons remain almost fully localized (hence excluded from the FS) even down to ~10 K.

Assuming completely localized 4f electrons, the Luttinger volume of the hole pockets (Bi 6p) should be the same as that of the electron pockets (Ce 5d), as required by the charge neutralization. However, we observe experimentally that the electron pocket near the $\overline{M}$ point expands considerably at lower temperatures, while the hole pockets near the $\overline{\Gamma}$ point remain unchanged (see momentum-distribution curves (MDCs) in Fig. 1(d,e) and energy-momentum cuts in Fig. 2(a,b)). Our detailed analysis indicates that the electron pocket expands mainly along the short axis (Fig. 1(e)), which results in a volume increase of ~ 40% at ~ 10 K compared to that at ~ 37 K, while no change can be detected for the hole pocket. This temperature dependence is confirmed by measurements from different samples (>10). The imbalanced change of hole and electron pockets implies that additional electrons (other than Bi-6p and Ce-5d) become involved in the FS at low temperature.

The temperature evolution of $k_F$ for the electron pocket is summarized in Fig. 2(c), where the top panel shows the raw data from one representative sample, demonstrating the onset temperature of ~ 35 K for the pocket expansion. To account for slight variations in $k_F$ for different samples and highlight the temperature evolution, we normalize the $k_F$ with respect to its high temperature value, i.e., $k_F$ (T) /$k_F$ (T > 35 K), and plot it at the bottom panel for four different samples. The result clearly shows that the pocket gradually expands with decreasing temperature and eventually reaches a plateau near 10 K. Since CeBi develops AFM order at ~25 K, the observed FS expansion might be related to the AFM transition. However, neither the FS maps in Fig. 1 nor the energy-momentum cuts in Fig. 2(a,b) show any sign of band folding at low temperature (more data can be found in [26]), indicating that the band folding effect from AFM ordering is very weak and does not play a significant role here. We also performed similar ARPES measurements for GdBi (Fig. 2(d)), an isostructural compound with AFM ordering at 25 K; the results show no temperature dependence for both the hole and electron pockets, in sharp contrast to CeBi (see [26] for additional data). In contrast to the 'unstable' $4f$ shell of CeBi with $4f$ occupancy $n_f \leq 1$, the half-filled one ($n_f = 7$) of GdBi is very 'stable'. Therefore, the observed expansion of the electron pocket in CeBi should most naturally be associated with the Ce-derived $4f$ electrons. This is also supported by the comparison of $\rho(T)$ in Fig. 2(e): while the resistivity of CeBi exhibits a clear Kondo-type upturn from high temperature to ~25 K and a sharp decrease below 25 K, the resistivity of GdBi shows only a tiny kink at the AFM transition, without any sign of Kondo-like behavior. For CeBi, the incremental resistivity, obtained after subtracting the result for LaBi ($\rho_i = \rho_{CeBi} - \rho_{LaBi}$), indeed follows the logarithmic temperature dependence characteristic for Kondo scattering (inset of Fig. 2(e)).

To better understand the Kondo effect and $c$-$f$ mixing, we performed resonant ARPES measurements at Ce $N$ edge ($4d \rightarrow 4f$) to enhance the spectral contribution from $4f$ electrons (Fig. 3(a)). We can clearly observe a sharp dispersionless $4f^0$ peak at ~-2.8 eV and the $4f^1$ peak at -0.3 eV, respectively. Here the labels, $4f^0$ and $4f^1$, refer to the electronic configuration after photoexcitation – therefore the $4f^0$ and $4f^1$ peak corresponds to the localized $4f$ electron (Ce$^{3+}$) and the many-body Kondo resonance (KR), respectively. The $4f^1$ peak at -0.3 eV is actually the spin-orbit split satellite (angular momentum $J=7/2$) of the KR, while the $J=5/2$ KR peak near $E_F$ is quite weak and almost absent (see Fig. 2(b) and Fig. 3(a)), similar to CeSb [27]. The presence of the $J=7/2$ peak implies that the Kondo process is indeed active, and the rather weak $J=5/2$ peak indicates that its $T_K$ must be very low [28,29]. The small $T_K$ is well expected for a low carrier density KL system, as the strength of the Kondo screening is strongly dependent on the density of conduction electrons [1].

Fig. 3(a) also shows the comparison of energy distribution curves (EDCs) near the hole and electron pockets at 16 K and 35 K, respectively. A very small decrease of the $4f^0$ peak intensity and a slight increase of the conduction bands near the $J=7/2$ $4f^1$ peak are seen at lower temperature. This trend becomes much clearer for soft X-ray resonant ARPES measurements at Ce $M$ edge (Fig. 3(b,c)), which is more bulk-sensitive due to higher energies. Below resonance (875 eV), the APRES spectra are dominated by the non-$f$ conduction bands, whose EDCs exhibit no temperature dependence (Fig. 3(c)). Upon cooling at resonance, e.g., at 880.6 eV and 881.2 eV, one can clearly observe a pronounced decrease of the $4f^0$ peak and a simultaneous intensity increase of the conduction bands from $E_F$ to ~ -1.5 eV. Additional data from another sample with a cooling and warming cycle is shown in [26]. Here the conduction bands are not resolved as clearly as those in Fig. 1 and Fig. 2, likely due to the limited resolution and low counts in soft X-ray ARPES. The pronounced temperature-dependent intensity transfer in Fig. 3(c) represents a direct evidence for the strong $c$-$f$ mixing at low temperature, which couples the wavefunctions of conduction and $f$ electrons, resulting in higher $f$ weight near $E_F$ (itinerant) and lower $f$ weight at $4f^0$ (localized) with decreasing temperature.

Our observation that both the spectral weight transfer in Fig. 3(c) and the FS expansion in Fig. 2(c) onset at $\simeq$ 35 K indicates that these two phenomena are intimately connected. According to Luttinger's theorem, the transfer of (a small amount of localized) $4f$ electrons into the conduction bands can lead to an expanded FS. Note that the FS change seen here is different from the small-to-large FS change observed in heavy fermion (HF) metals upon cooling

[30,31,32,33,34], during which a strong and coherent KR develops near $E_F$ and hybridizes with the conduction bands, resulting in an increased $k_F$. As no clear $J=5/2$ KR can be observed in CeBi [26], there must be a different cause for the FS expansion. In fact, our experimental results can be most naturally explained by a slight increase of Ce valence on cooling from ~ 35 to ~ 10 K, which we estimate to be ~ 0.01 based on the Luttinger count of the observed electron pocket. As stated in the introduction, such temperature dependence of the RE valence in KL systems appears to be quite natural [6,35,36]. In low carrier density systems, the many-body Kondo effect is largely diminished due to the stringent requirement for the conduction electron density and, therefore, the valence fluctuation originating in the substantial $c$-$f$ mixing can be resolved via the FS expansion. Such a small valence-induced change in FS is difficult to resolve in metallic KL systems, but it can be identified in low carrier density KL systems.

The observed intensity decrease of the $4f^0$ peak in Fig. 3(c) is much larger than the estimated valence increase of ~0.01 based on the Luttinger count of the electron pocket. This indicates that the observed spectral weight transfer in Fig. 3(c) should not be interpreted as being directly proportional to the actual number of transferred electrons. Instead, since the resonant ARPES intensity is proportional to the product of the electronic spectral function and the photoemission matrix element from the $f$ component (due to the resonance condition), the spectral weight transfer in Fig. 3(c) should contain a large contribution from the variation of matrix element, i.e., the reduced (increased) $f$ component in the wave function of $4f^0$ state (conduction bands) upon cooling.

The proposed $p$-$f$ mixing between the Bi-$6p$ and Ce-$4f$ states apparently does not change much within the measurement temperature window in CeBi [17], as evidenced by the weak $J=5/2$ KR and the lack of any resolvable change in the Bi $6p$ band. On the other hand, the large temperature-dependent expansion of the Ce 5$d$ band can be explained by a tiny increase of the Ce valence (~ 0.01), which could be checked by, e.g., high-resolution XAS at Ce $L$-edge [37]. The expansion of Ce 5$d$ pocket provides evidence for a strong covalent $d$-$f$ mixing among neighboring Ce ions, which contributes to the observed spectral transfer from the localized Ce-4$f$ states into itinerant Ce-5$d$ bands near $E_F$, i.e., the action of valence fluctuations. It is interesting to compare our results on CeBi with those of CeSb, where a recent ARPES study also revealed two different channels of interactions by the Ce-4$f$ electrons [27], i.e., the Kondo interaction (related to $p$-$f$ mixing) and magnetic exchange interaction (via $d$-$f$ mixing). We make the interesting observation that the $d$-$f$ mixing is obviously stronger in CeBi, compared to CeSb, with no sign of magnetic exchange splitting [27]. This difference is likely related to differences in the lattice constants and/or charge carrier densities.

To summarize, we have experimentally observed a large expansion (by ~ 40%) of the Ce-5$d$ pocket in CeBi at low temperature, while the Bi-6$p$ pockets remain unchanged. Resonant ARPES measurements at low temperature revealed pronounced spectral weight transfer from the 4$f^0$ state into the conduction bands near $E_F$, sharing similar temperature dependence with the FS expansion. The $J$=7/2 satellite of the KR could be clearly resolved, implying that the Kondo screening is operating, although a weak $J$=5/2 KR at $E_F$ indicates very low $T_K$. The observed FS expansion, distinct from the small-to-large FS change in metallic KL systems, can be explained by a small valence change of Ce (~ 0.01). These observations prove the existence of anisotropic $c$-$f$ hybridization. Our study demonstrates that weak valence fluctuations can be made visible in low-carrier density KL systems via significant thermally-driven changes in the FS, which can be important to understand relevant physical properties. The observed concurrence of real and virtual charge fluctuations as well as local-moment AFM order describes a state of quantum matter which has been enigmatic for years and deserves further scrutiny by future studies.

This work is supported by National Key R&D Program of the MOST of China (Grant No. 2017YFA0303100, 2016YFA0300203), National Science Foundation of China (No. 11674280, No. 11274006). We would like to thank Pengdong Wang, Dr. Caizhi Xu, Dr. Junzhang Ma for help in the synchrotron ARPES measurements. Y. L. thanks Profs. Qimiao Si and Stefan Kirchner for enlightening discussions. The synchrotron ARPES measurements was performed at BL13U beamline in Hefei national synchrotron radiation lab, SIS beamline at Swiss Light Source and 29-ID IEX beamline at Advanced Photon Source (APS). APS is supported by U.S. Department of Energy (DOE) Office of Science under Contract No. DE-AC02-06CH11357; additional support by National Science Foundation under Grant no. DMR-0703406 is also acknowledged.

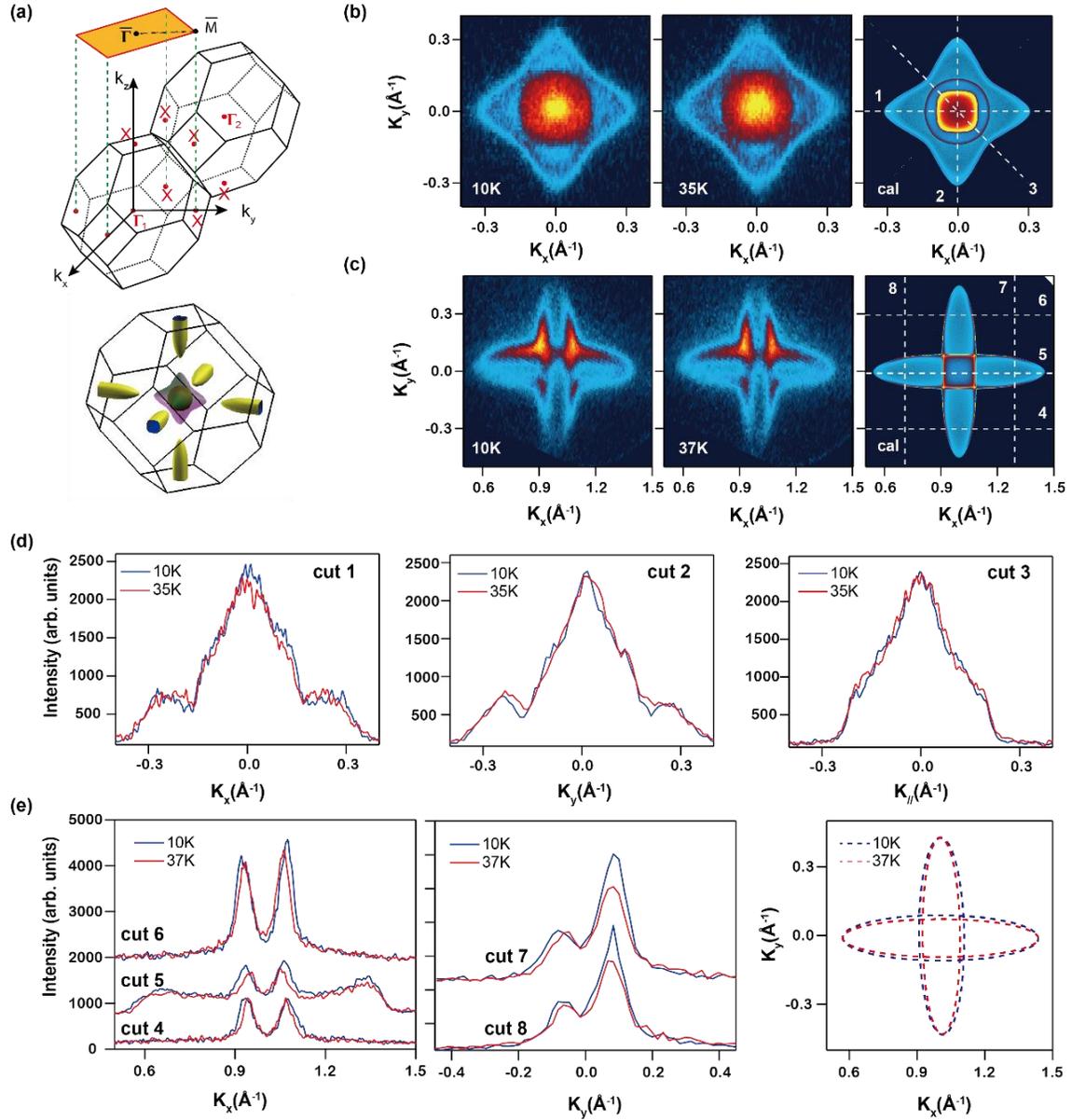

Fig. 1 (Color Online). Temperature-dependent FS of CeBi. (a) 3D and projected 2D BZs of CeBi (top), and the calculated 3D FS excluding Ce-4$f$ electrons (bottom). (b,c) Experimental FS at 10 K and ~ 37 K, in comparison with the DFT calculations, for both the hole pocket near the $\overline{\Gamma}$ point (b) and the electron pocket near the $\overline{M}$ point (c). (d,e) MDC cuts for the hole (d) and electron (e) pockets. The cut directions are indicated in (b,c). The rightmost panel in (e) shows the extracted contour of the electron pocket at low and high temperatures.

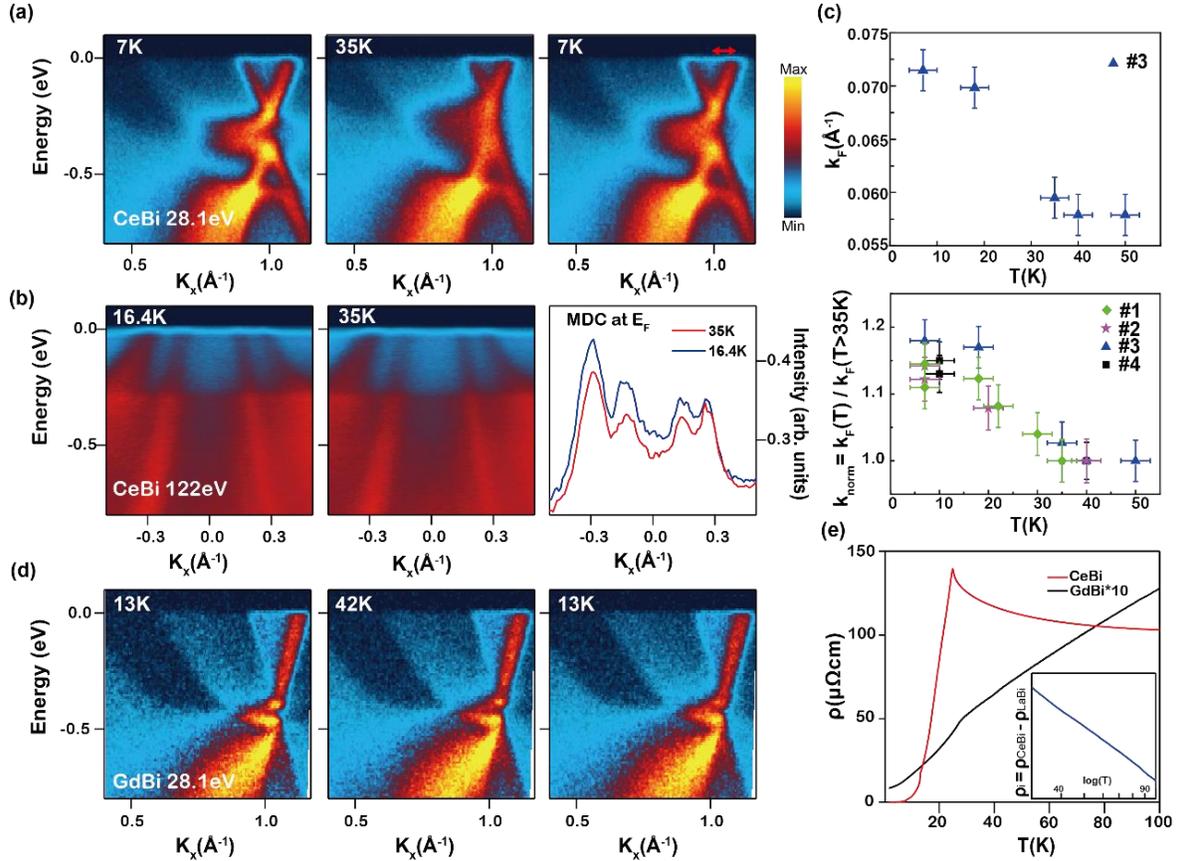

Fig. 2 (Color Online). Temperature evolution of the hole and electron pocket for CeBi, in comparison with GdBi. (a,b) temperature dependence of the energy-momentum cut for the electron (a) and hole (b) pocket. The data in (a) shows the evolution during a warming and cooling cycle from the same sample. The rightmost panel in (b) shows the MDC cuts at $E_F$; no change in $k_F$ can be observed. (c) Temperature evolution of $k_F$ for the electron pocket (indicated by a red arrow in (a)). The top panel displays the raw data from a representative sample (#3), showing the onset of pocket expansion at ~35 K. The bottom panel summarizes the normalized Fermi vector $k_{norm}=k_F(T)/k_F(T>35K)$, for different samples. (d) Temperature evolution of the electron pocket for GdBi, showing no discernible change. (e) Resistivity vs temperature for CeBi and GdBi. The inset is the incremental part of the resistivity of CeBi, $\rho_i=\rho_{CeBi}-\rho_{LaBi}$, plotted as a function of $\log T$.

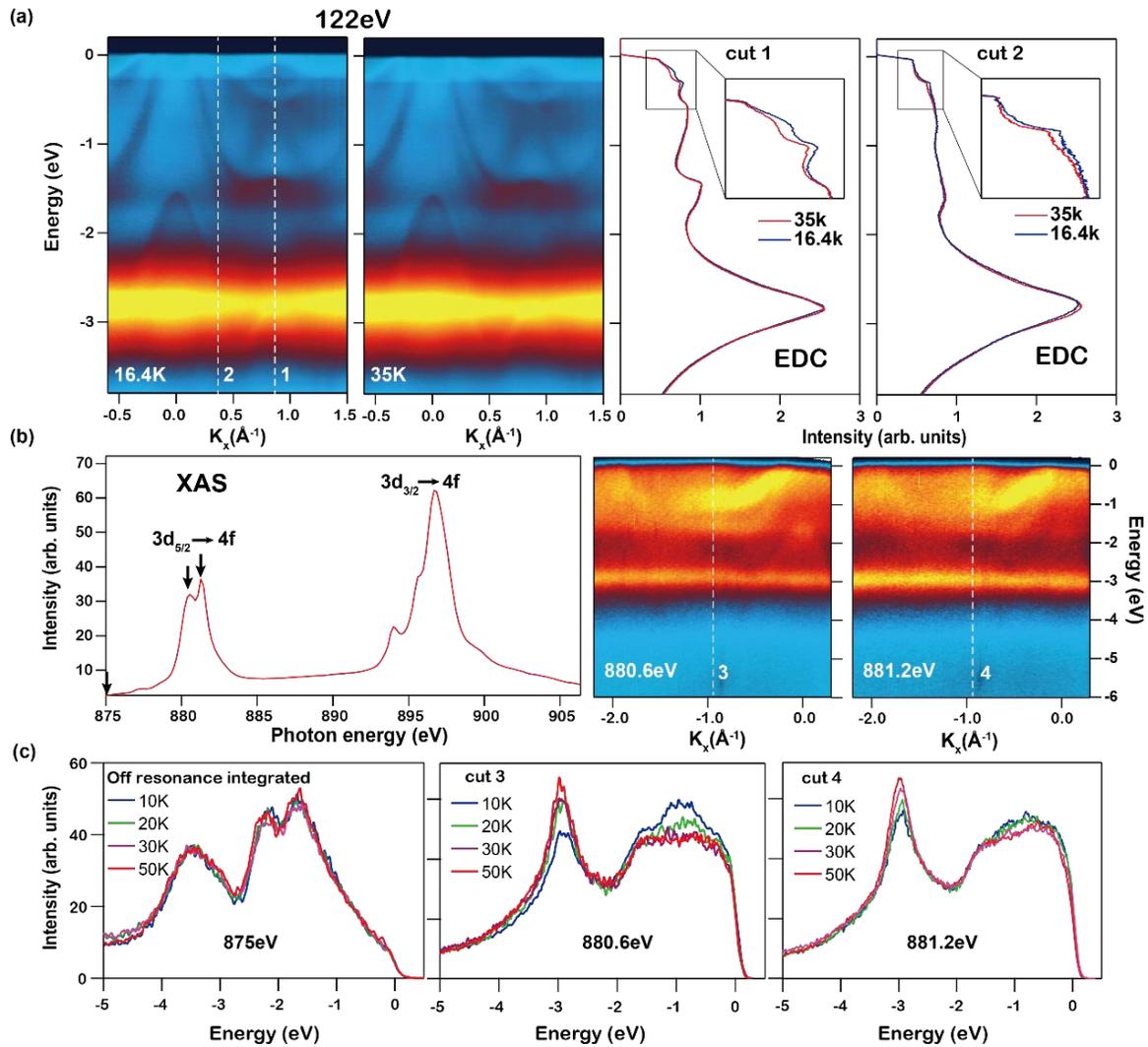

Fig. 3 (Color Online). Resonant ARPES results and temperature evolution. (a) Resonant ARPES results at Ce *N* edge (122 eV) at two representative temperatures. The EDC cuts at the hole and electron pockets are shown on the right panels, respectively. (b) Left: X-ray absorption spectroscopy (XAS) data near Ce *M* edge at 10 K. Right: Soft X-ray ARPES spectra taken at two representative photon energies (880.6 eV and 899.4 eV) at 10 K, indicated by black arrows in the XAS data. (c) Temperature dependence of EDCs at three representative photon energies (and momenta), indicated as white vertical lines in (b).